\documentclass[twocolumn,preprintnumbers,amsmath,amssymb]{revtex4-1}
\usepackage{graphicx}
\usepackage{dcolumn}
\usepackage{bm}
\usepackage[usenames,dvipsnames]{xcolor}
\usepackage[papersize={8.5in,11in}]{geometry}
\usepackage{float}

\begin{document}
%\preprint{ \color{Green} DRAFT v1.0 - \today}

\title{\Large {\color{blue}Controlled dripping from a grooved condensing plate}}

\author{M.Leonard}
\affiliation{GRASP, Institute of Physics B5a, University of Li\`ege, 4000 Li\`ege, Belgium}

\author{N.Vandewalle}
\affiliation{GRASP, Institute of Physics B5a, University of Li\`ege, 4000 Li\`ege, Belgium}

%%%%%%%%%%%%%%%%%%%%%%%%%%%%%  abstract

{\begin{abstract}
Condensed water on vertical surfaces ultimately leaves the substrate at the lower edge, where accumulated liquid detaches as drops. While droplet growth and surface transport have been extensively studied, this final release step remains poorly understood and largely uncontrolled. Yet this boundary event determines how and when condensed water is removed. We ask whether geometry can replace randomness as the governing mechanism of edge dripping.
By engraving vertical grooves upstream, we redirect water from surface flow into groove-guided drainage toward the boundary. This switch in transport mode changes how liquid accumulates and detaches at the edge. Using rapid forced condensation and high-resolution imaging, we systematically vary groove spacing $s$, aspect ratio $d/w$, and orientation. We then analyse how these geometric parameters influence the formation, stability, and spatial organization of droplets hanging below the edge. Smooth substrates exhibit irregular, impact-driven detachment. Grooved substrates produce localized and steady dripping points. When grooves converge, dripping occurs at fixed, geometry-defined locations.
For convergent designs, a simple condensation–capillarity model captures the dependence of the dripping period on the area of the drainage basin. Together, these results demonstrate that geometry alone can transform stochastic edge dripping into spatially organized and temporally regular release, with implications for dew harvesting, passive cooling, and millifluidic transport.
 \label{Abstract}
 \end{abstract}
}

\vskip 2 mm
\hrule 
\vskip -3 mm

%%%%%%%%%%%%%%%%%%%%%%%%%%%%% Intro
\maketitle

\section{Introduction}

Water flowing down vertical surfaces is a familiar sight in rainy climates, yet beneath this simplicity lies a remarkable diversity of morphologies. Depending on surface affinity and flow rate, water can appear as isolated droplets \cite{snoeijer_cornered_2007,le_grand-piteira_shape_2005}, narrow rivulets \cite{le_grand-piteira_meandering_2006, mertens_braiding_2004}, or continuous films \cite{schmuki_stability_1990, ghezzehei_constraints_2004,gabbard_thin_2023}. These behaviors have been widely explored on different surfaces, from wide flat panes \cite{rietz_dynamics_2017, le_grand-piteira_meandering_2006} to cones \cite{van_hulle_capillary_2021,chen_ultrafast_2018} and fibers \cite{quere_thin_1990,gilet_droplets_2010,gabbard_asymmetric_2021,zhang_droplets_2025,jambon-puillet_liquid_2019,darbois_texier_droplets_2015}, revealing how geometry and wetting properties shape drainage.

However, one region of these vertical surfaces has received far less attention: the lower edge, where the downward flow meets an abrupt discontinuity \cite{weyer_switching_2017,gilet_digital_2009}. At this boundary, water release depends on both the shape of the incoming liquid and the surface geometry \cite{lorenceau_capturing_2004, wang_numerical_2018,park_droplet_2017,jambon-puillet_gravito-capillary_2024}. A droplet may stick to the edge, a rivulet may cling and drip, and a film may shatter \cite{bremond_atomization_2007}. Despite its prevalence in both natural and engineered systems, the transition from surface transport to detachment at an edge remains poorly studied \cite{Lee_water_2012}.

This question becomes critical under condensation, where water forms continuously and must be removed efficiently \cite{muselli_dew_2009,rose_dropwise_2002,kim_dropwise_2011}. In dew collectors and heat exchangers, the fate of condensed water at the lower edge influences both retention and efficiency. Surface structuring offers a way to influence this process. Grooves are known to influence liquids in both static and dynamic ways. Statically, they pin contact lines \cite{quere_wetting_2008,luan_spontaneous_2021}, elongate droplets \cite{chen_anisotropy_2005,dai_hydrophilic_2018}, and absorb liquid through capillary suction \cite{bico_wetting_2002,seemann_wetting_2005, bamorovat_abadi_general_2020,van_hulle_effect_2023}; dynamically, they speed up transport along their length \cite{leonard_droplets_2023}, guiding droplets \cite{kern_twisted_2024,van_hulle_droplet_2024} or even stretching rivulets into thin films \cite{leonard_stretching_2025}. During condensation, grooves suppress gravitational shedding and draws the liquid into the grooves, allowing drainage within the texture rather than on the surface \cite{leonard_grooves_2025,bintein_grooves_2019}.

While grooves reorganize water transport along the vertical face, it remains unknown how this confined flow ultimately transitions into detachment at the lower edge. In particular, it is unclear how groove geometry shapes the moment and location at which condensed water leaves the substrate. Answering this question requires examining the lower edge directly, where groove-fed drainage meets gravity-driven detachment.

\section{Experimental Setup}

\begin{figure*}
    \centering
    \includegraphics[width=1\linewidth]{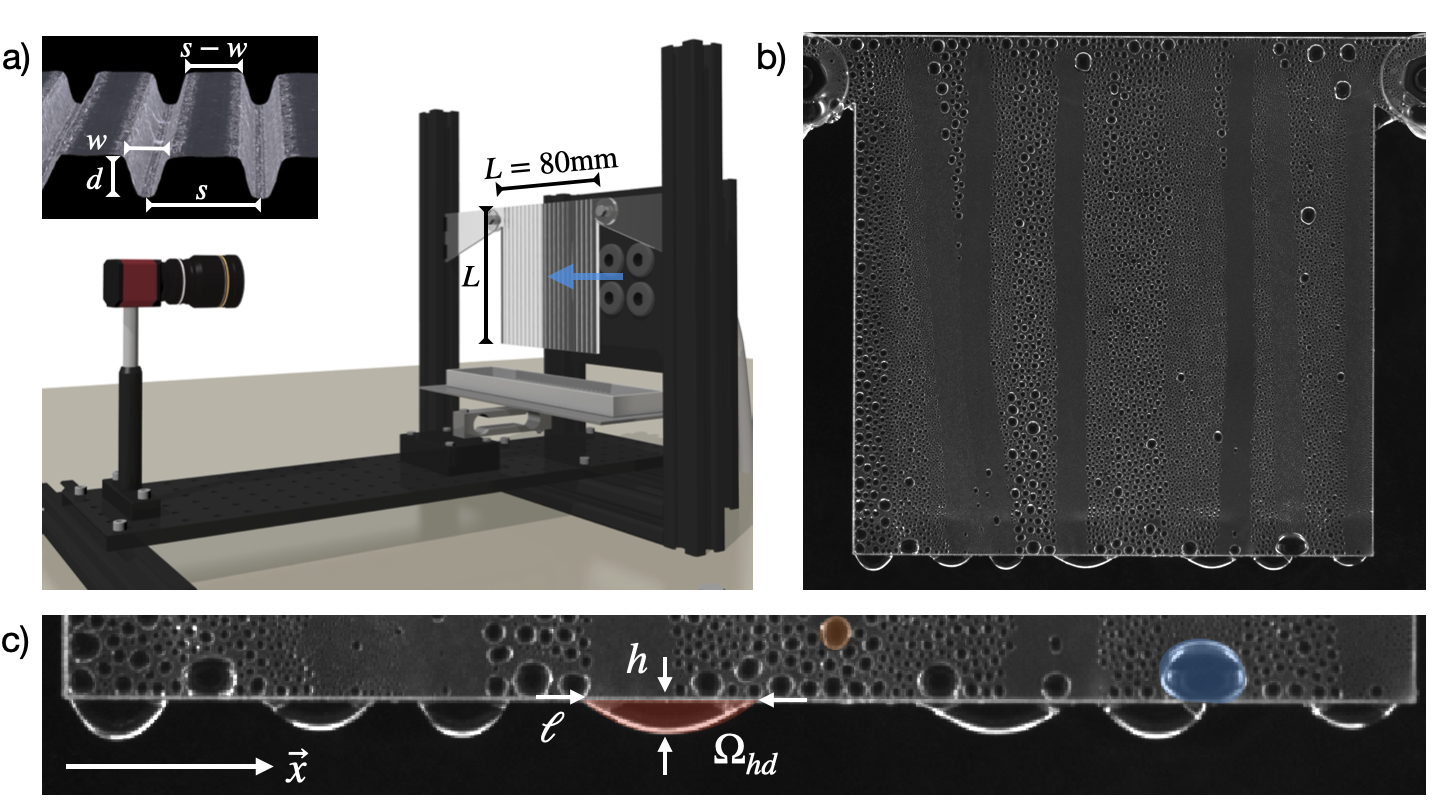}
    \caption{(a) Experimental setup. Warm, humid air is generated by bubbling compressed air through a heated water reservoir and directed toward the vertically mounted substrate via four nozzles (right). Condensation forms on a square acrylic plate of side length $L = 80.00\,\mathrm{mm}$, either smooth ($s = 80.00\,\mathrm{mm}$) or patterned with parallel vertical grooves. The grooves have a depth $d$ and a width $w$, while the spacing $s$ between grooves ranges from $0.30$ to $10.00\,\mathrm{mm}$. (b) Steady-state condensation on a smooth substrate showing the characteristic band structure formed by successive sweep drops sliding down the surface. (c)  Close-up view of the lower edge highlighting the three droplet types: (orange) droplets fully on the vertical face; (blue) flank droplets resting on the edge; (red) hanging droplets below the substrate.} 
    \label{fig: setup}
\end{figure*}

The traditional method of forced condensation involves cooling a surface by bringing it into contact with a thermal exchanger. Like condensation forming on a cold water bottle out of the fridge, water condenses when a surface is cooled below the dew point. While reliable, this technique is slow, typically yielding condensation rates around $6\,\rm{g/m^2\,h}$ \cite{trosseille_roughness-enhanced_2019,bintein_grooves_2019,jin_atmospheric_2017}. As a result, experiments often require several hours to complete.

Our approach takes the opposite route: instead of cooling the surface, we blow warm, humid air onto a room-temperature substrate, just like breathing on a cold window. Using this method, we achieve rates up to $900\,\rm{g/m^2\,h}$, 150 times faster, thereby reducing experiment time and enabling multiple tests across a wide range of parameters. Experiments are performed in a climate-controlled chamber at $T = 20.0 \pm 0.5^{\circ}\rm{C}$ and relative humidity $RH = 65 \pm 2\%$. To generate the humid airflow, we heat a water reservoir to $75 \pm 2^{\circ}\rm{C}$. Compressed air is injected through a diffuser at the bottom of the reservoir. As it rises, the air warms up through thermal exchange and becomes saturated with water vapor. This warm, moist air exits through four nozzles located on the reservoir lid and flows perpendicularly toward the vertically suspended substrate at a velocity $v < 1\,\rm{m/s}$ (see Figure~\ref{fig: setup}). These velocities are comparable to those occurring in natural dew formation environments \cite{muselli_dew_2009,jacobs_passive_2008}.

The substrate reaches a temperature of $T = 45.0 \pm 1.5^{\circ}\rm{C}$ within about 100 seconds. At this temperature, liquid water has a density $\rho = 992\,\rm{kg/m^3}$ and surface tension $\sigma = 69 \, 10^{-3}\,\rm{N/m}$ \cite{gittens_variation_1969}, corresponding to a capillary length of $\lambda = \sqrt{\sigma / \rho g} = 2.67\,\rm{mm}$. The substrate is a square acrylic plate (TroGlass Clear) with thickness $e = 3 \,\rm{mm}$ and side length $L = 80\,\rm{mm}$. Its wetting properties are characterized by a static advancing contact angle $\theta_A = 78.0 \pm 4.5^\circ$ and a receding angle $\theta_R = 48.0 \pm 4.7^\circ$, yielding an average contact angle of $\theta = 63.0\pm 4.6^\circ$. 

Surface structuring is done with a laser cutter (Trotec Speedy 100), producing grooves with spacing $s$ ranging from $0.30$ to $10.00\,\rm{mm}$, with depth $d$ and width $w$ (specified when appropriate and measured using optical microscopy (Keyence VHX), see inset of Figure \ref{fig: setup}). A smooth surface ($s = 80.00\,\rm{mm}$) serves as the reference. Each experiment lasts 50 minutes (3000s) and is repeated three times to ensure reproducibility.

\section{Results}
Condensation on a smooth vertical surface proceeds in a well-known sequence. Nucleation begins at material imperfections, which act as preferential sites for droplet formation~\cite{lavielle_memory_2023}. Droplets grow first by vapor adsorption and then by coalescence with neighbors. Once a droplet reaches the critical radius $R_c \approx 1\,\rm{mm}$ \cite{leonard_grooves_2025}, its weight exceeds the surface retention force, and it starts sliding downward~\cite{extrand_retention_1990,gao_how_2018}. These sliding droplets collect smaller ones along their path~\cite{rose_dropwise_2002}, rapidly increasing in size. For this reason, we call these drops sweep drops. As they descend, droplets may become unstable and break up into smaller ones due to the Rayleigh–Plateau instability~\cite{podgorski_corners_2001,le_grand-piteira_shape_2005}, limiting their volume and leaving behind a narrow, nearly dry trail of width approximately $2\lambda$ wide. This dry wake provides a clean region for renewed nucleation, setting up a continuous condensation–drainage cycle. Because droplets near the top of the surface experience less frequent sweeping, they have more time to grow and are more likely to initiate a shedding event. As a result, drainage is managed progressively, droplet by droplet, from the top part of the surface, as illustrated by the vertical stripes in Figure~\ref{fig: setup}(b). To understand how sweep drops drainage translates into dripping, we now focus on the lower edge of the substrate, shown in Figure~\ref{fig: setup}(c). We identify three distinct droplet types: droplets entirely on the vertical surface (orange), flank droplets resting on the horizontal edge (blue), and hanging droplets located below the substrate (red).

\begin{figure*}
\centering
\begin{minipage}[t]{0.48\textwidth}
    \centering
    \includegraphics[width=\textwidth]{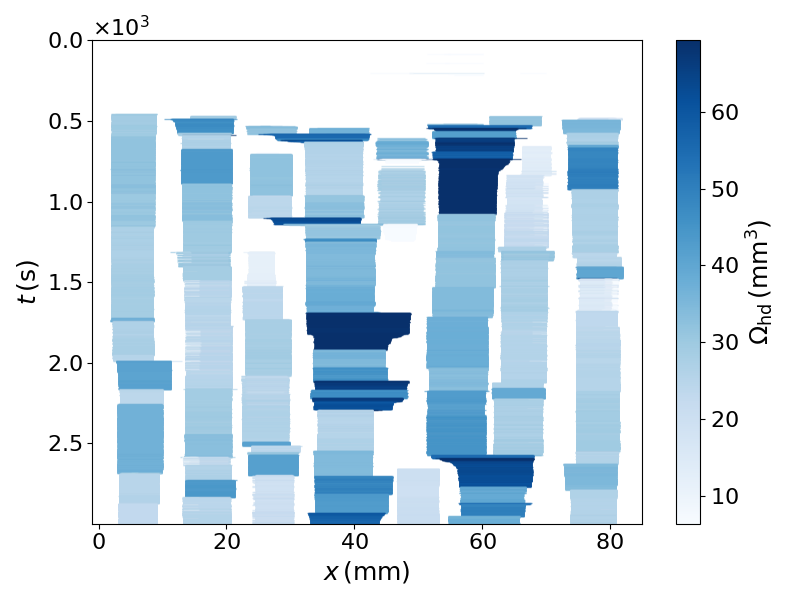}\\
    (a) $s=80.00\,\mathrm{mm}$
\end{minipage}
\begin{minipage}[t]{0.48\textwidth}
    \centering
    \includegraphics[width=\textwidth]{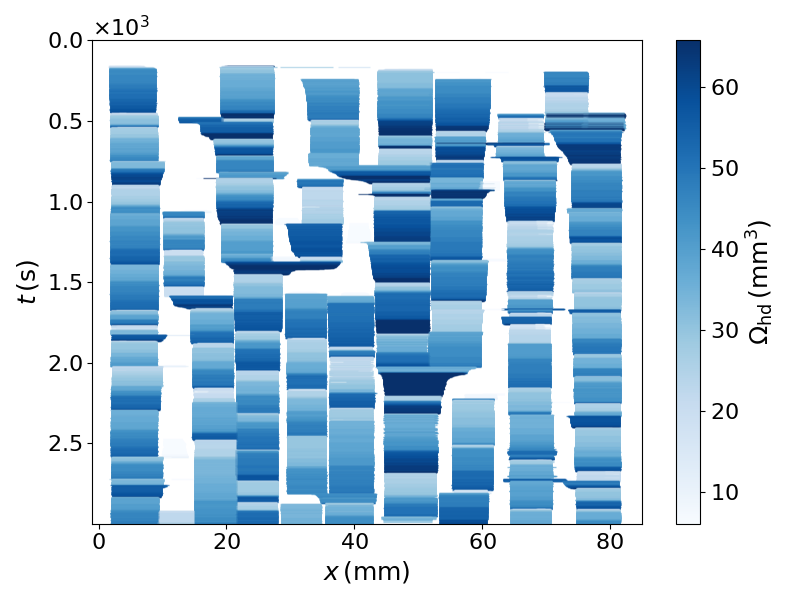}\\
    (b) $s=1.25\,\mathrm{mm}$
\end{minipage}
\begin{minipage}[t]{0.48\textwidth}
    \centering
    \includegraphics[width=\textwidth]{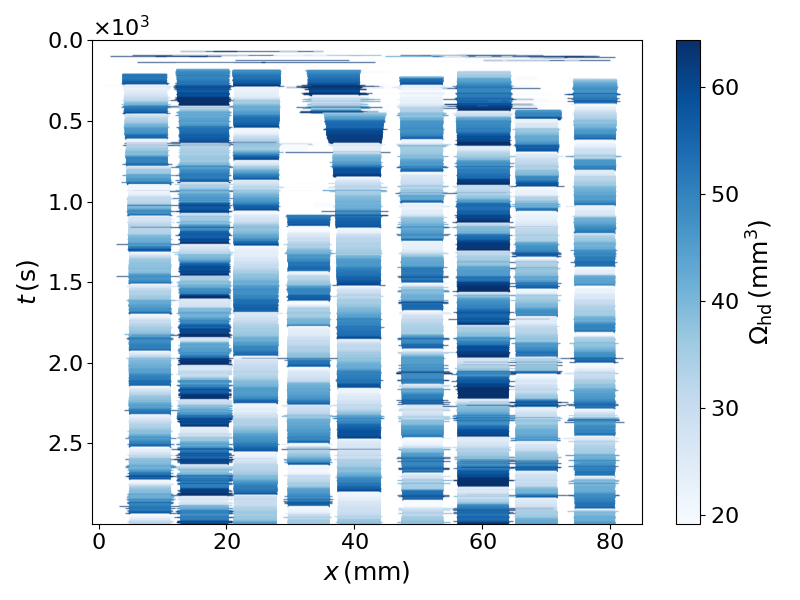}\\
    (c) $s=0.50\,\mathrm{mm}$
\end{minipage}
\begin{minipage}[t]{0.48\textwidth}
    \centering
    \includegraphics[width=\textwidth]{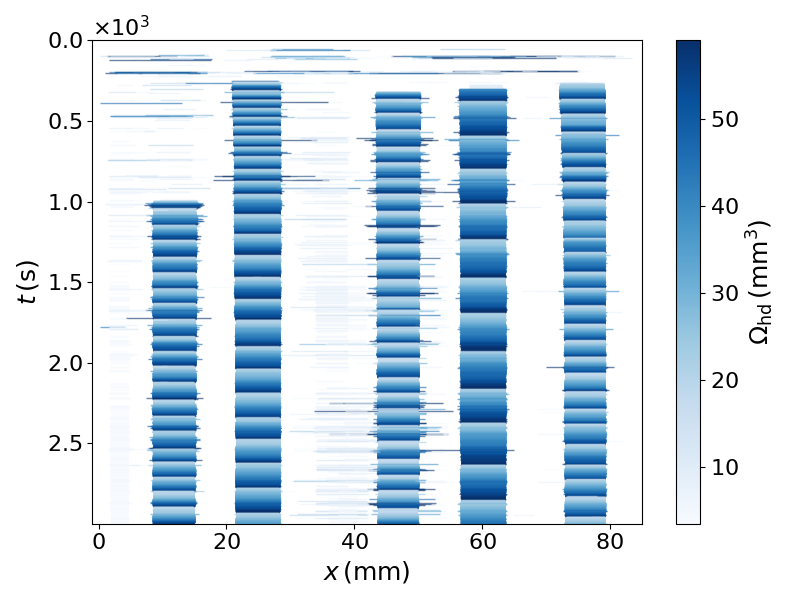}\\
    (d) $s=0.30\,\mathrm{mm}$
\end{minipage}
\caption{Spatio-temporal evolution of hanging droplets below the lower edge for decreasing groove spacing:
(a) $s = 80.00\,\mathrm{mm}$, (b) $1.25\,\mathrm{mm}$, (c) $0.50\,\mathrm{mm}$, (d) $0.30\,\mathrm{mm}$.
Each bar represents one droplet; its length corresponds to droplet width $\ell$ and its colour to volume $\Omega_{hd}$. The vertical axis is time $t$, while the horizontal axis corresponds to position $x$ along the edge.
At large spacing (a), hanging droplets appear after a latency of $\sim500\,\mathrm{s}$ and form irregular, drifting bands produced by sporadic impacts of sliding droplets.
As grooves are introduced (b–c), bands become thinner, straighter, and more numerous, reflecting increased positional stability and smoother volume evolution.
For the most closely spaced grooves (d), the number of hanging droplets decreases, but their positions and volume variation periods become highly regular. The dripping pattern transitions from intermittent and impact-driven to structured and quasi-periodic as groove spacing decreases.}
\label{fig: hanging_spacing}
\end{figure*}

We now examine how hanging droplets evolve during condensation. Figure~\ref{fig: hanging_spacing}(a) presents the spatio-temporal evolution of hanging drops on a smooth substrate ($s = 80.00\,\mathrm{mm}$). The horizontal axis indicates droplet position $x$ along the edge, the vertical one indicates time $t$ evolution, from top to bottom, and the color refers to drop volume $\Omega_{hd}$. We estimate the latter as a half-ellipsoid
\begin{equation}
\Omega_{hd} = \tfrac{1}{6}\pi \ell h e,
\end{equation}
with drop width $\ell$, height $h$, and thickness $e$ approximated to the substrate thickness ($3\,\mathrm{mm}$).
At first glance, the figure shows a set of colored bands extending vertically across the plot. The pattern itself is composed of narrow horizontal bars, each representing one frame of observation (recorded every second), with its length proportional to the droplet width $\ell$ and its color scaled by the volume $\Omega_{hd}$. Hanging droplets appear after a latency of about $500\,\mathrm{s}$, once sweep drops from the vertical face reach the lower edge. Each band is thus a sequence of droplets replacing one another as they detach. These bands reveal that droplets repeatedly form at the same location along the edge, but these positions may shift gradually over time. Their color changes abruptly, revealing sudden variations in droplet volume due to coalescence or impact with a sweep drop. Bands merge as neighboring hanging droplets coalesce, and reappear when there is sufficient room for a sweep or flank droplet to convert into a new hanging one. The outcome is a self-organized yet unpredictable dripping pattern, governed by the intermittent arrival of sweep drops from the vertical face.

To explore how grooves alter this behavior, we gradually decrease their spacing $s$. Groove geometry is $d=211\pm 7 \,\rm{\mu m}$ and $w=206\pm 5 \,\rm{\mu m}$ yielding a aspect ratio $d/w=1.02$. When the first grooves are introduced ($s = 1.25\,\mathrm{mm}$), the dripping pattern changes noticeably. The bands become more numerous, thinner, and straighter, showing greater stability. Their color transitions are smoother, suggesting more gradual filling. Interactions between neighboring bands are more frequent, even if droplets now appear in a more continuous and evenly spaced way. At narrower spacing ($s = 0.50\,\mathrm{mm}$), the pattern becomes strikingly regular. Bands are thin, vertical, and closely packed, while color oscillations are strong and periodic, marking steady cycles of filling and drainage. The width of each band remains constant in time, and droplet positions no longer shift. At the smallest spacing ($s = 0.30\,\mathrm{mm}$), the number of hanging droplets strikingly decreases and their positions become highly stable. Bands are perfectly vertical and widely spaced, with rapid and repetitive color oscillations. Large inactive regions remain between them, and subtle correlations appear: the droplet located at $x \approx 25\,\mathrm{mm}$ at $t\approx1000\,s$ on Figure \ref{fig: hanging_spacing}(d) slows its dripping when a new droplet begins to fill at $x \approx 15\,\mathrm{mm}$. The dripping pattern settles into a stable rhythm, reflecting the geometric constraint imposed by the underlying texture.

\begin{figure}
\centering
    \begin{minipage}[t]{0.48\textwidth}
        \centering
        \includegraphics[width=1\linewidth]{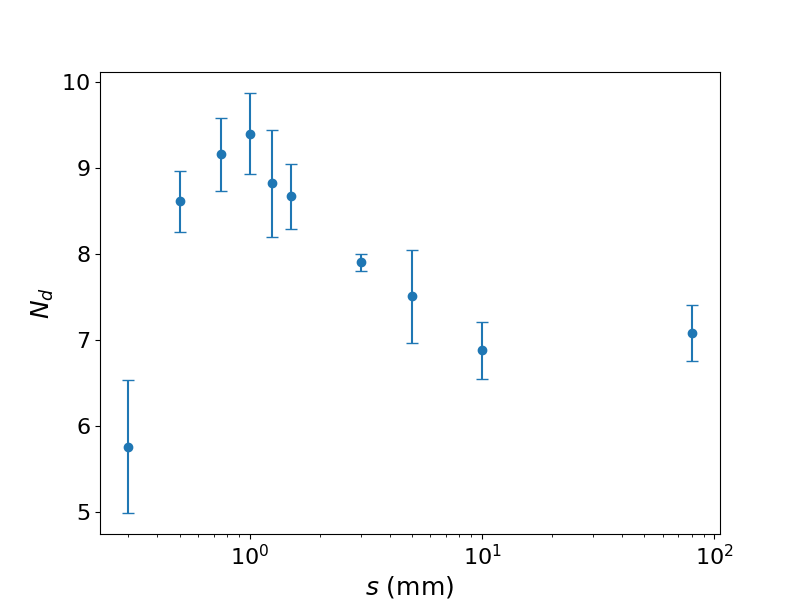}\\
        (a) 
    \end{minipage}
    \begin{minipage}[t]{0.48\textwidth}
        \centering
        \includegraphics[width=1\linewidth]{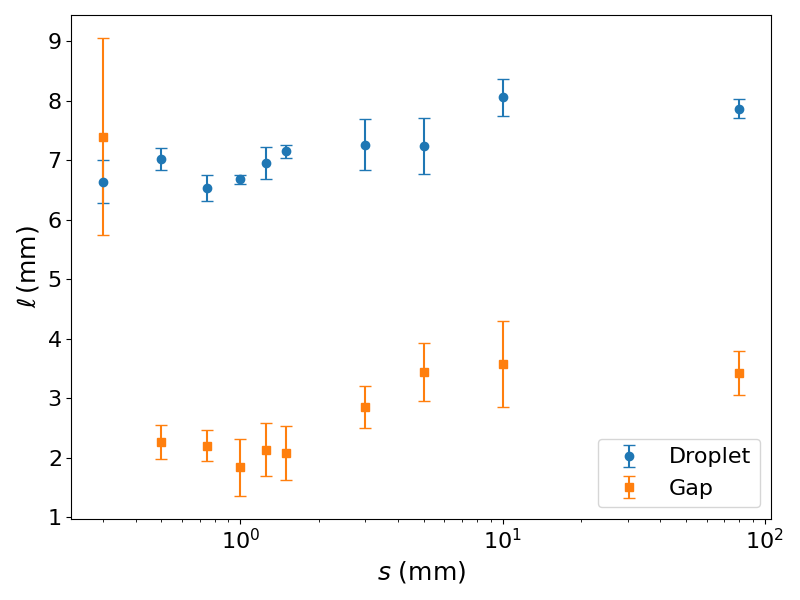}\\
        (b) 
    \end{minipage}
\caption{Evolution of hanging droplets in the steady regime ($t > 1000\,\mathrm{s}$).
(a) Mean number of hanging droplets $N_d$ as a function of groove spacing $s$. Two regimes appear: for large and intermediate spacings ($s > 1\,\rm{mm}$), the number of droplets increases as grooves are added; for narrow spacing ($s < 1\,\rm{mm}$), it decreases sharply as grooves become denser. This behavior closely mirrors the water-retention trend observed on the vertical face, with a transition near $s \approx R_c$.
(b) Mean droplet width $\ell$ and mean inter-drop gap in the same steady regime. Droplet width decreases continuously as spacing narrows, from about $8\,\mathrm{mm}$ on the smooth surface to $6.5\,\mathrm{mm}$ on the most structured ones ($s < 1\,\mathrm{mm}$), corresponding roughly to twice the capillary length ($2\lambda$). The mean gap follows a similar trend for $s > R_c$ but shifted with a value around $\lambda$. Then it rises again for $s < R_c$, reaching values comparable to droplet size $2\lambda$.
Together, these measurements reveal a transition from intermittent, gravity-driven dripping to a geometrically constrained, periodic regime as groove spacing decreases.}
\label{fig: hanging_N_w}
\end{figure}

Figure \ref{fig: hanging_N_w} quantifies how groove spacing $s$ influences the dripping regime. In panel (a), the mean number of hanging droplets $N_d$ exhibits two distinct trends. For large and intermediate spacings ($s > 1\,\rm{mm}$), decreasing groove spacing increases the number of droplets; but once spacing falls below $1\,\rm{mm}$, $N_d$ drops abruptly. Panel (b) shows that a systematic diminution in droplet size accompanies this shift. The mean width $\ell$ decreases from about $8\,\mathrm{mm}$ on the smooth surface to $6.5\,\mathrm{mm}$ for the most structured samples ($s < 1\,\mathrm{mm}$), remaining close to twice the capillary length ($2\lambda$). The mean inter-drop gap behaves similarly for $s > 1\,\mathrm{mm}$, hovering near $\lambda$, but widens again once the grooves are densely packed, reaching values comparable to droplet size. Taken together, these measurements reveal a clear shift in dripping behavior with groove spacing: the system evolves from a random, gravity-fed pattern to a more gradual and stable one as the surface becomes more structured. The transition occurs around $s \approx R_c$, coinciding with the spacing where water transport along the vertical face changes from sweep drops drainage to groove-guided \cite{leonard_grooves_2025}. This correspondence suggests that the reorganization of dripping at the edge is not an isolated effect, but rather the continuation of a deeper transformation in how water moves across the entire surface, a transition whose mechanism becomes fully apparent when the grooves are most densely packed.

\section{Discussion}

\begin{figure*}
\centering
\includegraphics[width=0.8\textwidth]{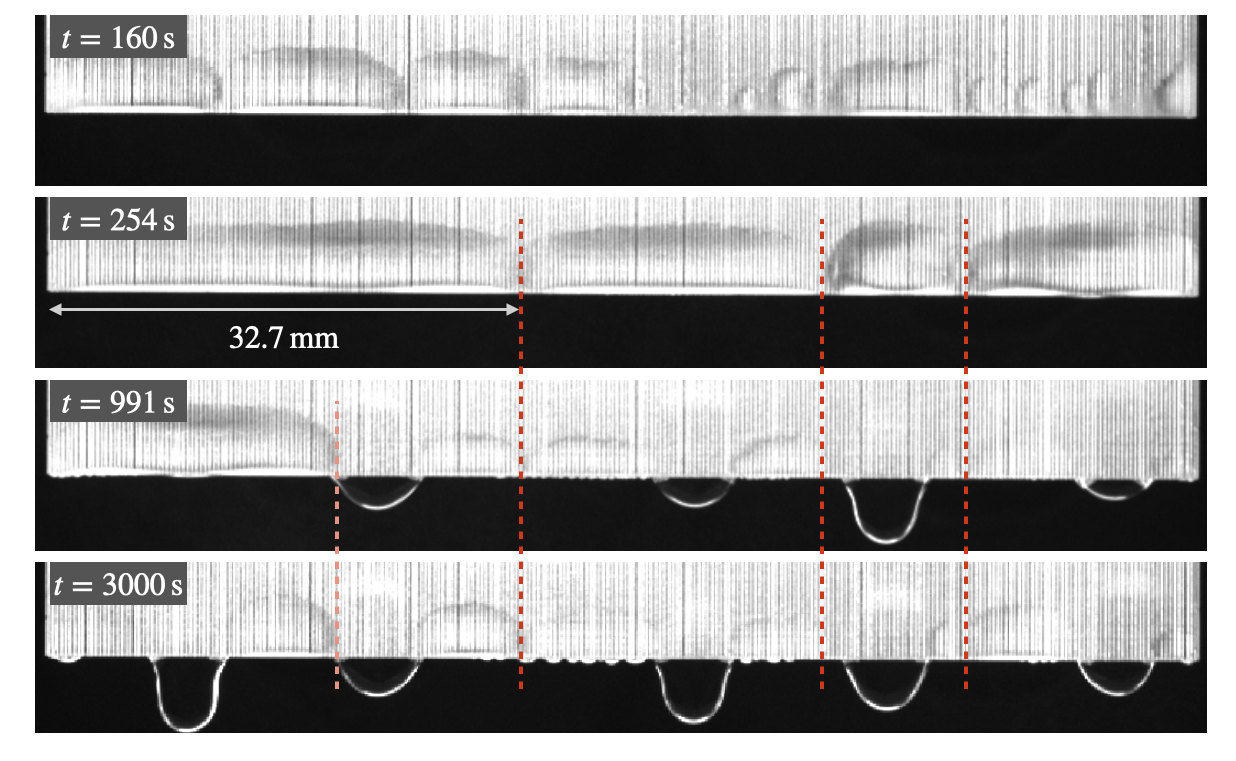}
\caption{
    Time evolution of flank and hanging droplets on a densely grooved substrate ($s = 0.30\,\mathrm{mm}$, edge length $80\,\mathrm{mm}$).  
    Snapshots at four key moments illustrate how groove-fed drainage gives rise to a stable dripping pattern at the lower edge.  
    ($t = 160$–$254\,\mathrm{s}$): rapid formation and lateral merging of groove-fed flank droplets, some spanning over $30\,\mathrm{mm}$ in width;  
    ($t = 991\,\mathrm{s}$): local detachment begins as the last large flank droplet is about to break, marking the final stage before the system settles;  
    ($t = 3000\,\mathrm{s}$): the configuration remains unchanged long after the last breakup, reflecting the persistent steady state established shortly after $t = 991\,\mathrm{s}$.  
    Red and rose dashed lines outline "fragmentation basins", zones where successive generations of flank droplets have broken up and stabilized. The groove network pins the boundaries of these basins, locking the spatial organization of the dripping pattern in place.
}
\label{fig: hd_03_time_illu}
\end{figure*}

To understand how groove structuring organizes dripping, we focus on the most densely patterned substrate ($s = 0.30\,\mathrm{mm}$), illustrated in Figure~\ref{fig: hd_03_time_illu}, with its spatiotemporal evolution shown in Figure~\ref{fig: hanging_spacing}(d). At this spacing, sweep drop drainage is completely suppressed, allowing us to isolate the sequence of events that transforms groove-fed flow into a stable dripping pattern at the lower edge.

At early times ($t = 160$–$254\,\mathrm{s}$), a first generation of flank droplets appears along the lower part of the vertical face. These droplets are fed through the grooves, which rapidly transport water downward, faster than it would under gravity alone. As they grow, the droplets often merge laterally, forming extended flank droplets that span several grooves. Some reach widths up to $33\,\mathrm{mm}$, over ten times the capillary length $\lambda$. Such large droplets are unusual: based on the Bond number, which compares gravitational and capillary forces,
\begin{equation}
    \mathrm{Bo} = \frac{\rho g \ell^2}{\sigma} = \frac{\ell^2}{\lambda^2} \approx 150,
\end{equation}
they should detach under their own weight. Instead, they remain pinned, stabilized by the grooves acting as capillary anchors \cite{bico_wetting_2002,seemann_wetting_2011}. Yet, the slightly curved lower contours of these droplets suggest that gravity is beginning to overcome capillary pinning (at $t=254\,\rm{s}$). A few seconds later, the heaviest flank droplets destabilize locally. The first hanging droplets appear directly below the points where the curved edges of the flank droplets begin to sag. Detachment is not necessarily symmetric; breakups occur at irregular locations along the droplet base, likely triggered by small imperfections at the substrate's lower edge. Each pendant droplet thus traces its origin to a previous and specific flank droplet, marking the coupling between the two types. Moreover, no flank droplet is ever observed directly above a hanging one. Once a groove supplies a hanging droplet, its capillary connection prevents the formation of a new flank droplet in the same vertical line. 

After several minutes ($t = 991\,\mathrm{s}$), the last large flank droplet is about to break. This moment marks the final stage before the system settles into a stable, organized configuration. All the droplets that previously dominated the lower edge have fragmented, sometimes in multiple stages, into an alternating sequence of flank and hanging droplets. Through this cascade, each droplet progressively reaches equilibrium, resulting in a self-organized state that emerges from the balance between gravity and capillarity.

By the end of the experiment ($t = 3000\,\mathrm{s}$), more than $2000\,\rm{s}$ after the last major breakup, the droplet arrangement remains unchanged. The same flank droplets persist, continuously replenished through the grooves and periodically discharging into adjacent hanging droplets. The spatial distribution established after the first breakups persists, as highlighted by the red and rose dashed lines in Figure~\ref{fig: hd_03_time_illu}, which delineate "fragmentation basins", zones where successive generations of flank droplets have broken up and stabilized.

In the closed-spacing configuration, stability arises from the interplay between grooves and flank droplets. The grooves suppress sweep-drop drainage and anchor flank droplets to the vertical face, shielding them from destabilizing impacts. These anchored flank droplets then capture and channel condensed water toward the hanging ones, establishing preferential feeding sites.
Together, they create a groove-fed drainage network that stabilizes hanging droplets and renders dripping localized and regular. However, the exact release position ultimately reflects the stochastic breakup of the large initial flank droplets.

\begin{figure*}
\centering
\includegraphics[width=0.9\textwidth]{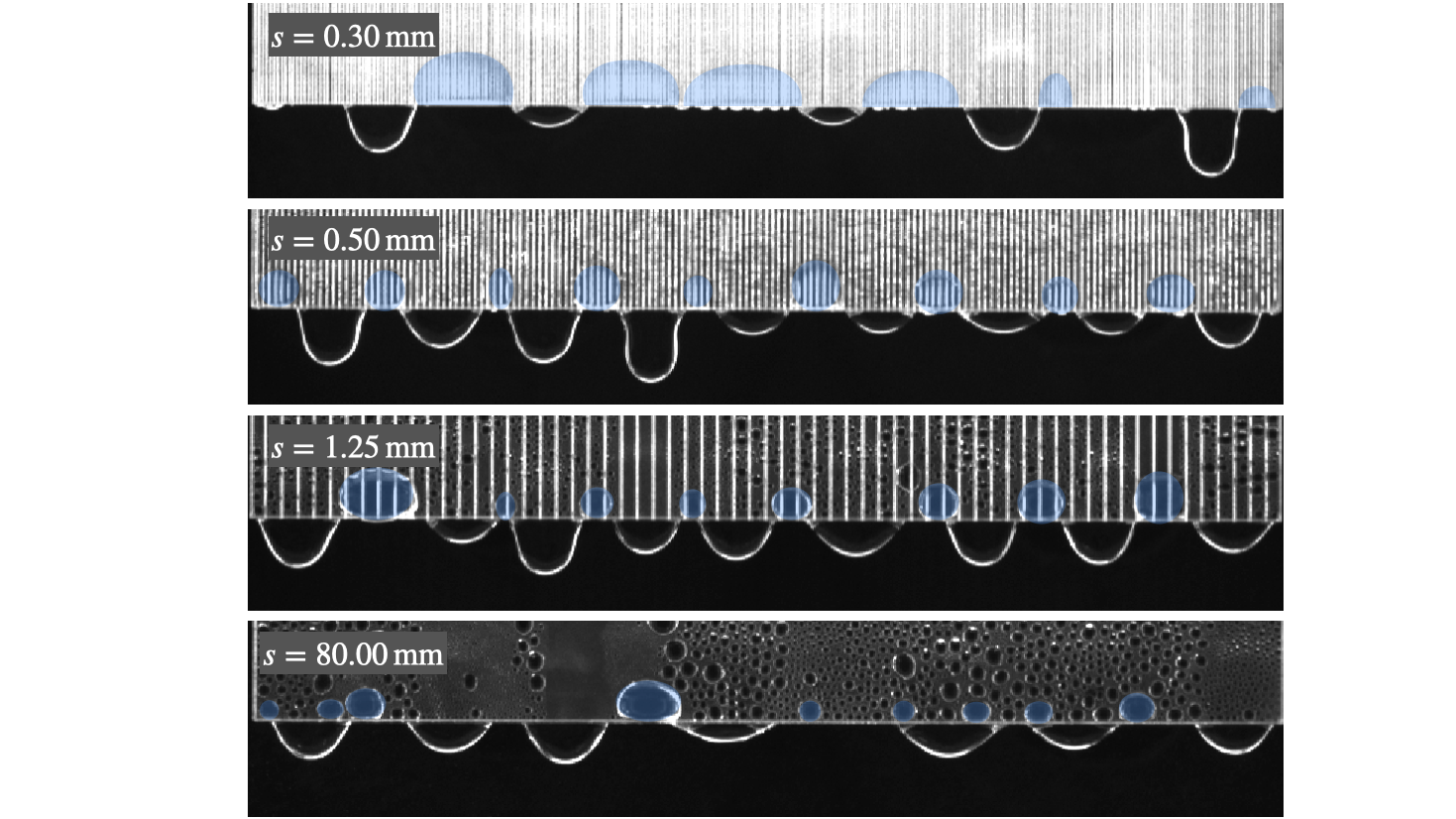}
\caption{Snapshot of the bottom of the sample at $t = 2500\,\mathrm{s}$ for increasing groove spacing: $s = 0.30-0.50-1.25-80.00\,\mathrm{mm}$.
Side-anchored (flank) droplets are highlighted in blue to aid visualization.
As groove spacing increases, droplet organization evolves from an alternating arrangement of hanging and flank droplets to random deposition on the smooth substrate, reflecting the disappearance of groove-mediated capillary-driven drainage.}
\label{fig: sample_hd}
\end{figure*}

Having identified how dense grooving produces a stable dripping configuration, we now ask the reverse question: what happens as this structure gradually disappears? In other words, how does groove density control the stability of the dripping regime?
To address this, we compare substrates with progressively larger groove spacing ($s = 0.30$–$0.50$–$1.25$–$80.00\,\mathrm{mm}$) under steady condensation in Figure \ref{fig: sample_hd}.

At the smallest spacing $s = 0.30\,\mathrm{mm}$, groove density is highest and sweep drop drainage is entirely suppressed. As described earlier, flank droplets form first and are fed directly by the grooves. These elongated droplets are strongly stabilized by capillary anchoring: the dense network of grooves resists dewetting and holds them in place despite their large volume. In turn, these stable flank droplets act as anchors for the hanging ones below, maintaining both their position and rhythm.

When spacing increases to $s = 0.50\,\mathrm{mm}$, the groove density decreases but remains sufficient to prevent sweep droplet from forming. Water still arrives exclusively through the grooves, yet the anchoring weakens because each droplet now spans fewer grooves. As a result, flank droplets become smaller and more circular. Since the lower boundary is entirely occupied, either by flank droplets or by hanging ones, any reduction in flank coverage leaves room for new hanging droplets to form. In other words, weakening the anchoring of flank droplets directly promotes the appearance of hanging ones. This explains the rising branch of the $N_d(s)$ curve for $s < 1\,\rm{mm}$ in Figure \ref{fig: hanging_N_w}.

At $s = 1.25\,\mathrm{mm}$, groove anchoring weakens further, and sweep drop drainage starts to compete again. Water now reaches the bottom edge intermittently, sometimes through grooves, sometimes through occasional sweep drops. Because the grooves are farther apart, each drainage event delivers larger, more sudden volumes of liquid. These abrupt inflows destabilize the flank droplets, which may either detach to form new hanging droplets or be swept away entirely. As a result, flank droplets lose their stabilizing role, and the dripping pattern becomes irregular. This transitional behavior near $s \approx R_c$ marks the point where sweep drops begins to overpower capillary organization.

On the smooth surface ($s = 80.00\,\mathrm{mm}$), no grooves remain, and capillary anchoring disappears entirely. Drainage is now dominated by fast sweep drops descending at about $100\,\mathrm{mm/s}$. Flank droplets can still form transiently at the corner, resting on the geometric discontinuity of the edge. Still, they are short-lived, with width $\leq\lambda$ and easily dislodged by the impact of descending droplets. The two populations, flank and hanging droplets, are thus completely decoupled and often appear superposed in the same region of the sample.
Yet, despite repeated high-speed impacts, preferential hanging sites persist along the edge. Individual droplets detach, but new ones repeatedly form at nearly the same locations: it is the site, not the droplet, that remains stable (see Figure \ref{fig: hanging_spacing}(a)). This resilience reflects the balance between inertia and capillarity, with surface tension remaining comparable to the impact forces. A simple estimate of the Weber number,
\begin{equation}
    \rm{We} = \frac{\rho r v^2}{\sigma} \approx 0.5
\end{equation}
confirms it. When a sweep drop hits a hanging one, it removes part of the volume of the hanging droplet, but not the whole droplet. In many cases, the reverse happens: the hanging droplet pulls the sweep one toward itself and absorbs it. As a result, hanging droplets maintain a relatively stable position over time. Their mean width, about $2\lambda$, and spacing, about $\lambda$, define a gravity–capillarity equilibrium that arises even without grooves. This explains the descending branch of $N_d(s)$ for large $s$, where dripping is sparse, uncoordinated, and governed purely by the balance between inertia and surface tension.

\begin{figure*}
\centering
\begin{minipage}[t]{0.48\textwidth}
    \centering
    \includegraphics[width=\textwidth]{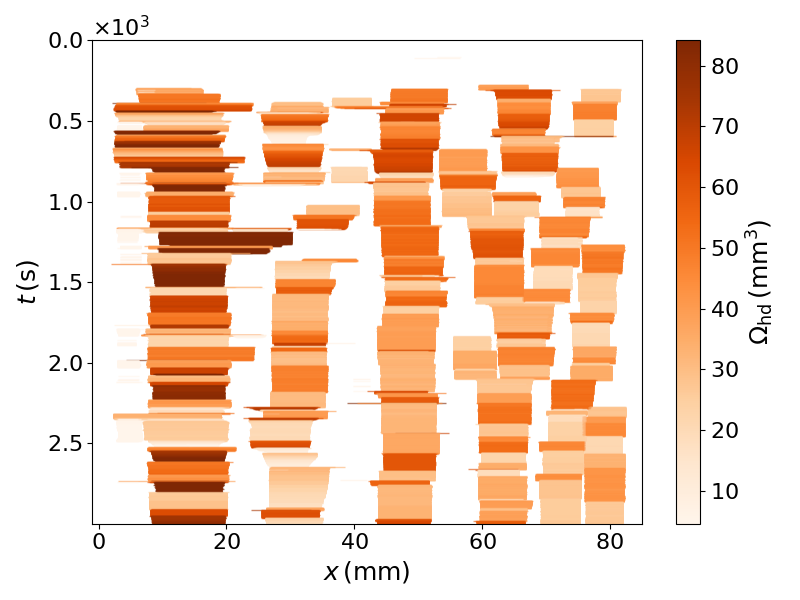}\\
    (a) $d/w = 0.20$
\end{minipage}
\begin{minipage}[t]{0.48\textwidth}
    \centering
    \includegraphics[width=\textwidth]{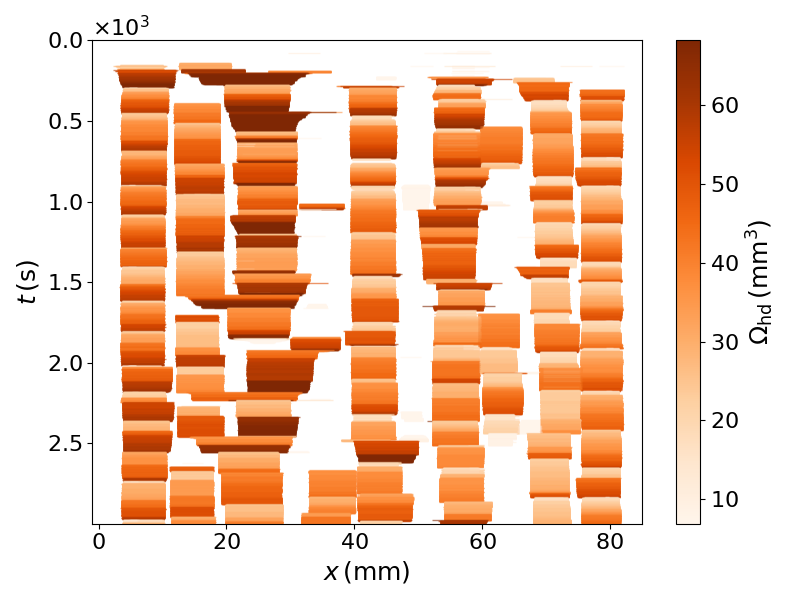}\\
    (b) $d/w = 0.28$
\end{minipage}
\begin{minipage}[t]{0.48\textwidth}
    \centering
    \includegraphics[width=\textwidth]{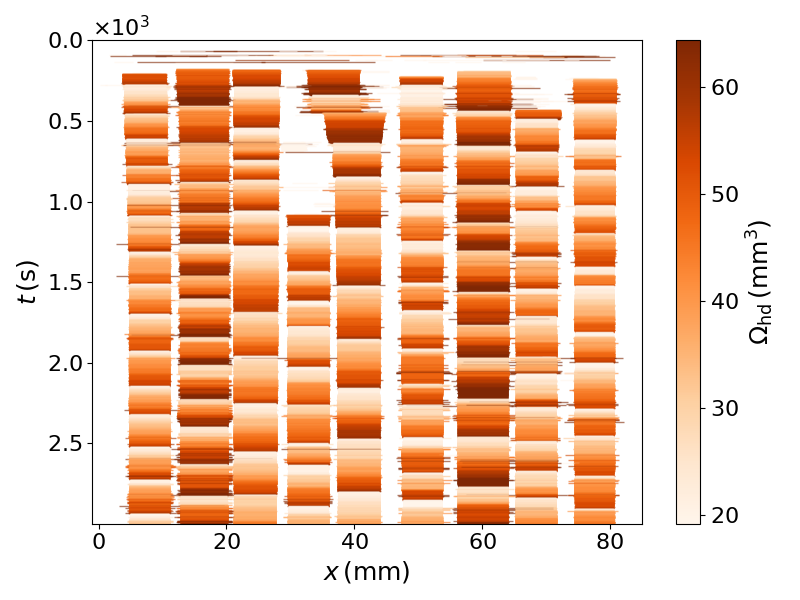}\\
    (c) $d/w = 0.95$
\end{minipage}
\begin{minipage}[t]{0.48\textwidth}
    \centering
    \includegraphics[width=\textwidth]{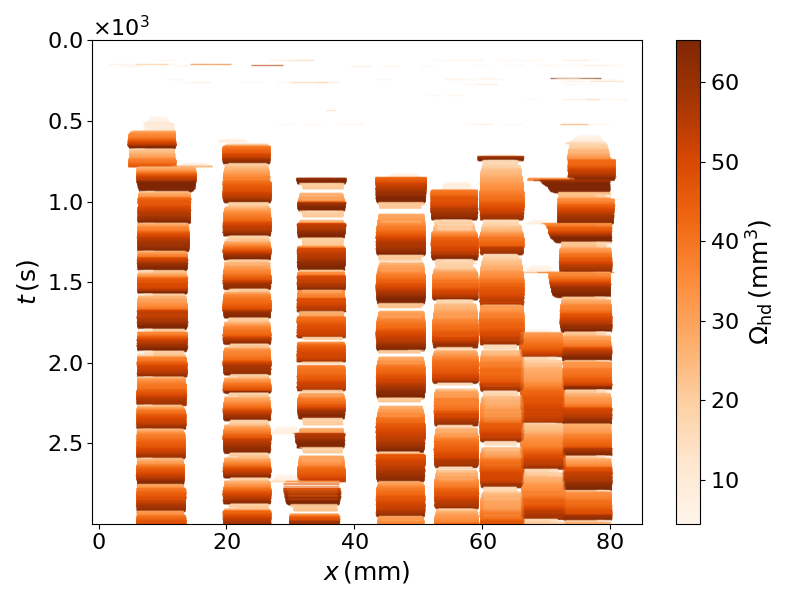}\\
    (d) $d/w = 3.01$
\end{minipage}
\caption{Spatio-temporal evolution of hanging droplets below the lower edge for increasing groove aspect ratio at fixed spacing $s = 0.5\,\mathrm{mm}$:
(a) $d/w = 0.20$, (b) $0.28$, (c) $0.95$, and (d) $3.12$.
Each bar represents a droplet; its length corresponds to the droplet's width, and its color corresponds to the droplet's volume.
As grooves deepen, the dynamics follow the same progression observed when decreasing spacing in Figure \ref{fig: hanging_spacing}:
Shallow grooves ($d/w < 0.3$) yield sparse, irregular dripping dominated by gravity, whereas deep grooves ($d/w > 1$) promote capillary confinement and produce regular, quasi-periodic dripping.
The transition illustrates how increasing groove depth enhances anchoring strength, shifting the system from a gravity-driven to a capillary-organized regime.}
\label{fig: hanging_aspect_ratio}
\end{figure*}

So far, we have varied the anchor density by changing the groove spacing. We now turn to their strength: each groove's geometry, through its depth and width, determines its ability to hold or release water, and, in turn, reshapes the dynamics of dripping. To investigate this effect, we compare grooves of increasing aspect ratio $d/w$: $0.20$ ($d = 28\,\mathrm{\mu m}$, $w = 140\,\mathrm{\mu m}$), $0.28$ ($d = 58\,\mathrm{\mu m}$, $w = 210\,\mathrm{\mu m}$), $0.95$ ($d = 204\,\mathrm{\mu m}$, $w = 215\,\mathrm{\mu m}$), and $3.01$ ($d = 783\,\mathrm{\mu m}$, $w = 260\,\mathrm{\mu m}$).

The spatio-temporal patterns obtained when varying the groove aspect ratio (Figure \ref{fig: hanging_aspect_ratio}) mirror those observed for spacing (Figure \ref{fig: hanging_spacing}). Shallow grooves behave similarly to sparsely spaced ones: they provide weak anchorage while promoting sweep drops, resulting in irregular, gravity-dominated dripping. Deep grooves, by contrast, act like closely packed ones, strengthening capillary anchors and suppressing gravitational drainage. It results in stable and periodic dripping. In both cases, the system moves through the same sequence of regimes as the balance between capillarity and gravity shifts. Groove spacing thus controls the density of anchors, while aspect ratio sets their strength, two complementary parameters that together define the architecture of the dripping pattern.

\section{Convergent grooves}

We have demonstrated that parallel grooves can affect the geometry and dynamics of dripping, but not its precise location. In other words, grooves organize drainage, yet the dripping points remain statistically distributed along the lower edge. To localize the dripping more precisely, water must not only be guided downward through capillary flow, but also directed toward predetermined positions. This can be achieved by replacing the parallel pattern with a convergent one.

\begin{figure}
\centering
\begin{minipage}[t]{0.48\textwidth}
    \centering
    \includegraphics[width=0.7\textwidth]{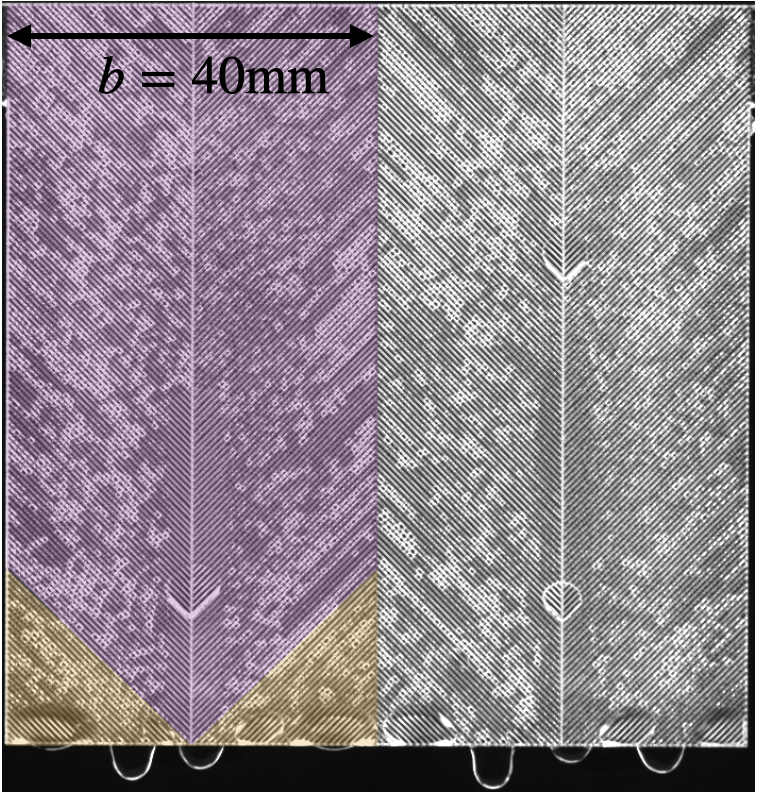}\\
    (a) 
\end{minipage}
\begin{minipage}[t]{0.48\textwidth}
    \centering
    \includegraphics[width=\textwidth]{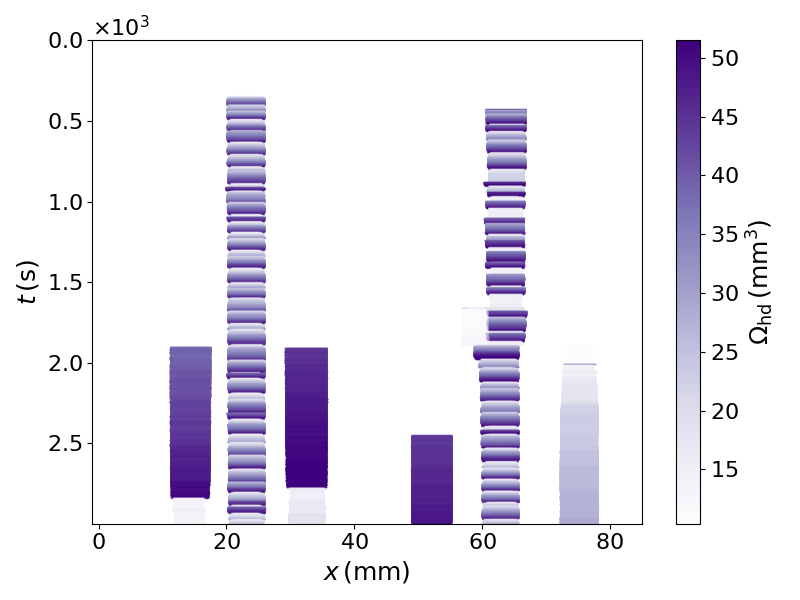}\\
    (b)
\end{minipage}
\caption{(a) Convergent groove pattern during steady condensation. One $b=40\,\mathrm{mm}$-wide pattern (highlighted in purple) is repeated twice across the $80\,\mathrm{mm}$ substrate. Each unit features a central vertical groove and multiple secondary grooves inclined at $45^{\circ}$ and spaced by $0.50\,\mathrm{mm}$, converging toward the center. The two convergent basins collect water from both sides and channel it into the vertical grooves at $x = 20$ and $60\,\mathrm{mm}$. Triangular regions at the base of each pattern (highlighted in yellow) remain unconnected to the main basin.
(b) Spatio-temporal diagram of hanging droplets below the convergent substrate. Two dominant vertical bands correspond to the dripping points located beneath the vertical grooves. Rapid color oscillations indicate intense and periodic drainage cycles, while faint, slowly varying bands outside these regions correspond to peripheral droplets. Nearly all condensed water exits through the two central outlets, demonstrating precise localization of dripping.}
\label{fig: hanging_convergent}
\end{figure}

Figure~\ref{fig: hanging_convergent}(a) shows a convergent groove pattern composed of repeating $40\,\mathrm{mm}$-wide units (one highlighted in purple for clarity). The substrate, $80\,\mathrm{mm}$ wide, therefore contains two such units. Each pattern consists of a central vertical groove flanked by multiple secondary grooves inclined at $45^{\circ}$, spaced by $0.50\,\mathrm{mm}$ and of aspect ratio $d/w=0.53$ ($d=138\,\rm{\mu m}$ and $w=260\,\rm{\mu m}$) which ensures efficient capillary drainage. On each side, the inclined grooves converge toward the central one, of the same geometry as the secondary ones, forming two drainage basins centered at $x = 20$ and $60\,\mathrm{mm}$.

Within each basin, condensed water first joins one of the inclined grooves, which channels it toward the central vertical groove. There, the accumulated flow sometimes exceeds the drainage capacity \cite{leonard_stretching_2025}, producing droplets astride the central groove, visible in the picture. As a result, water reaches the bottom edge in two ways: continuously through the groove channels, and intermittently via these overflowing droplets. In both cases, the flow converges to a single point on the edge. The position of dripping is thus no longer determined by random breakups of flank droplets, but by the imposed groove geometry itself.

Outside the main basins, the inclined grooves leave two small triangular regions at the base of each pattern, unconnected to the vertical groove (highlighted in yellow in Figure~\ref{fig: hanging_convergent}(a)). Water condensing in these peripheral areas is directed straight to the lower edge, where it accumulates as small flank droplets that can occasionally evolve into hanging ones. However, because these regions collect only a minor fraction of the total condensed water, the resulting droplets grow very slowly and contribute little to the overall drainage.

The spatio-temporal diagram in Figure~\ref{fig: hanging_convergent}(b) confirms the effectiveness of the convergent design. Two dominant bands appear, each corresponding to one vertical groove. These bands exhibit rapid, regular color oscillations, reflecting high-frequency filling and discharge cycles, and remain fixed in position within one capillary length ($\lambda$) of the groove axis. A few secondary bands also appear outside these main regions, but their color variations are much slower, indicating that the associated droplets evolve weakly. These correspond to the peripheral zones described earlier. Thus, while isolated droplets may still appear elsewhere, nearly all the condensed water is channeled to the two main outlets. The dripping is fully localized.

\begin{figure}
    \centering
    \begin{minipage}[t]{0.48\textwidth}
        \centering
        \includegraphics[width=1\linewidth]{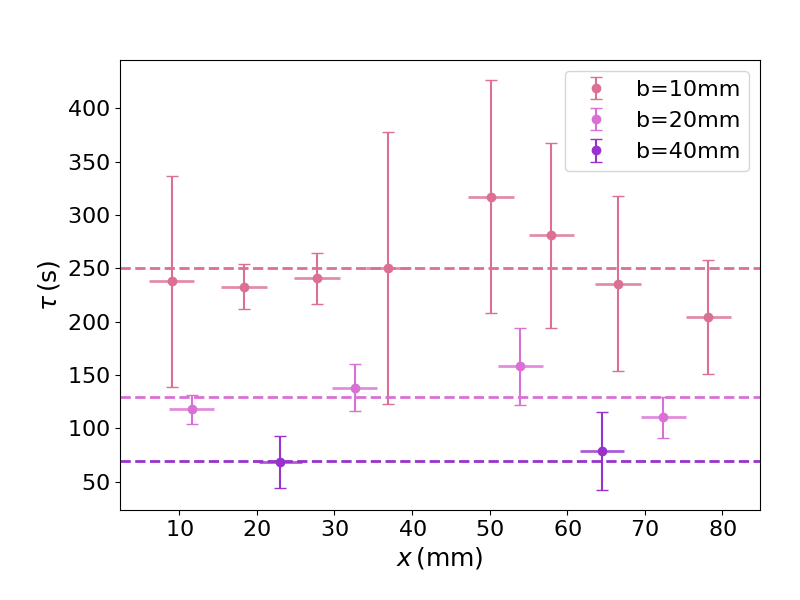}\\
    \end{minipage}
    \caption{Average dripping period and position of the active bands for three basin widths ($40$, $20$, and $10\,\mathrm{mm}$). Each point corresponds to a stable dripping source identified in the spatio-temporal diagrams. The number and position of active sources match the vertical grooves defining each basin, within mean droplet width $\ell$ which is indicated by the horizontal line.
    Dotted lines show the prediction of the simple condensation–capillarity model for each basin width $b$ (Equation \ref{eq:tau}). The agreement confirms that each basin behaves as a capillary attractor, with its surface determining the period of water release.}
    \label{fig: period_convergent}
\end{figure}

Finally, we examine how the basin width $b$ influences the dripping period. The period $\tau$ is extracted from the temporal evolution of the droplet volume within each stable band. Figure~\ref{fig: period_convergent} shows the average dripping period $\tau$ and the corresponding edge position $x$ for three basin widths, $b = 40$, $20$, and $10,\mathrm{mm}$ (one experiment per configuration).
For a given pattern, all basins exhibit nearly identical dripping periods. Moreover, the number of active dripping sources matches exactly the number of basins, and their positions coincide, within $\ell/2$, with the vertical grooves defining each basin. This confirms that each convergent basin acts as an independent drainage unit that fixes both the location and timing of release.
The slightly higher median values and broader error bars are consistent with the $1\,\mathrm{fps}$ temporal resolution. It limits the accurate capture of the fastest dripping events, as well as small volume changes before and after detachment. This effect is most visible in the central region ($x \in [20, 60]\,\mathrm{mm}$), where marginally higher condensation rates shorten the dripping period.

To rationalize this trend, we consider the time $\tau$ required for condensation to generate the mass $m_{hd}$ of a single hanging droplet ready to detach. The effective condensing area $S$ of one basin, of width $b$ and height $L$, excluding the triangular regions at its base, is
\begin{equation}
S = b(L - b/4).
\end{equation}
Given a measured condensation rate $c = 0.208\,\mathrm{g\,s^{-1}\,m^{-2}}$, the total accumulated mass over time $\tau$ is $m_{hd}=cS\tau$. We estimate the characteristic droplet mass through a capillary–gravity balance, 
\begin{equation}
m_{hd} = \frac{2(\ell + e)\sigma}{g},
\end{equation}
where $\ell$ and $e$ denote the droplet width and substrate thickness, respectively.
Equating both expressions yields the expected dripping period,
\begin{equation}
\tau = \alpha\,\frac{m_{hd}}{c\,b(L - b/4)},
\label{eq:tau}
\end{equation}
where $\alpha = 0.25$ is a single empirical prefactor determined from nine independent experiments (three for each basin width) and applied uniformly to all configurations. On the one hand, this factor accounts for the fact that only a fraction of each hanging droplet's mass actually detaches during each cycle. On the other hand, it reconciles the discrepancy between the simple capillary–gravity mass estimation of $m\approx 0.15\,\mathrm{g}$ and the measured droplet mass of $m \approx 0.05\,\mathrm{g}$ obtained using a balance placed beneath the condensing plate.

Convergent grooves, therefore, provide a robust and straightforward strategy to control both the position and the period of dripping. By designing the geometry of the drainage basins, one can dictate both where water leaves the substrate and how often it does so.

\section{Conclusion}

This study demonstrates that the groove texturing governs not only the drainage of condensed water along a vertical substrate but also its detachment at the lower edge. By combining controlled condensation with high-resolution imaging, we identified how groove spacing, depth, and orientation dictate the emergence, stability, and coordination of hanging droplets.

When grooves are widely spaced, sweep drop impacts dominate drainage, leading to irregular dripping. As spacing decreases or grooves deepen, capillary forces at the groove edges stabilize flank droplets that feed the hanging ones below, transforming random dripping into a periodic and organized regime. The groove spacing thus sets the density of pinning sites, while the aspect ratio tunes their strength. Finally, convergent groove patterns enable complete control over the dripping position and period, turning the lower edge into an engineered array of capillary outlets.

Together, these results reveal a continuous transition from gravity-driven drainage to capillarity-organized release, demonstrating that edge detachment can be synchronized and localized solely through geometry. Beyond condensation, this framework provides a generic route to designing surfaces that precisely manage liquid release, relevant to dew and fog harvesting, passive cooling, and millifluidic transport. Future work could explore dynamic or hierarchical textures to couple droplet formation, flow guidance, and edge emission within a single integrated surface.

%%%%%%%%%%%%%%%%%%%%
%\vskip{1cm}
\section*{Acknowledgments} 
Valsem Industries SAS financially supported this work. The authors would like to thank Prof. D. Beysens for fruitful discussions, Prof. E. Parmentier for the microscope measurements, Mr M. Melard and Mr S. Rondia for their contributions to the experimental setup, and Mr Y. Corbisier for the experimental diagram. Prof. N. Vandewalle thanks the Francqui Foundation for financial support.

%%%%%%%%%%%%%%%%%%%%
\vskip 0.6 cm
\section*{Author information} 
\subsection*{Corresponding Author}

\subsection*{Author contributions} 

\section*{Note}
The authors declare no competing financial interest.

%%%END OF MAIN TEXT%%%

%%%%%%%%%%%%%%%%%%%%
%\bibliographystyle{achemso.bst}
\bibliography{MyLibrary2}

\end{document}